\documentstyle[pra,aps,graphicx,twocolumn]{revtex}
\begin{document}
\title{Momentum state engineering and control in Bose-Einstein condensates}
\author{Sierk P\"otting$^{1,2,3}$\thanks{email: sierk.poetting@optics.arizona.edu},
        Marcus Cramer$^{1,4}$,
        and Pierre Meystre$^{1}$}
\address{$^1$Optical Sciences Center, University of Arizona, Tucson, Arizona 85721}
\address{$^2$Max--Planck--Institut f\"ur Quantenoptik, 85748 Garching, Germany}
\address{$^3$Sektion Physik, Universit\"at M\"unchen, 80333 M\"unchen, Germany}
\address{$^4$Fachbereich Physik der Philipps--Universit\"at, 35032 Marburg, Germany}

\date{\today}

\maketitle

\begin{abstract}

We demonstrate theoretically the use of genetic learning
algorithms to coherently control the dynamics of a Bose--Einstein
condensate. We consider specifically the situation of a condensate
in an optical lattice formed by two counterpropagating laser
beams. The frequency detuning between the lasers acts as a control
parameter that can be used to precisely manipulate the condensate
even in the presence of a significant mean-field energy. We
illustrate this procedure in the coherent acceleration of a
condensate and in the preparation of a superposition of prescribed
relative phase.
\end{abstract}


\section{Introduction}
\label{sec:intro}

Atomic Bose-Einstein condensates (BEC) are now generated almost
routinely in the laboratory, and are resulting in numerous
applications in fundamental and applied science. A key element of
this research is the ability to coherently engineer and control
the state of the condensate and its dynamics. This is for example
the case in the generation of atomic solitons
\cite{Burg99,Dobr99,Dens00}, which requires that a precise phase
be imprinted on the condensate; atomic four-wave mixing, which
involves the splitting of a condensate into three momentum groups
\cite{Deng99}; the mixing of optical and matter waves
\cite{Inou99a,Inou99b,Law98,Moor99,Meys01}, which again requires
the splitting of a condensate; mode-locked atom lasers
\cite{Ande98}, where a condensate wave function is spatially
modulated by a periodic optical potential, etc. It can be
expected that the preparation of condensate states of increasing
sophistication will be required in future applications, including
atom lithography \cite{Drod97,John98} and atom holography
\cite{Zoba99}.

The coherent manipulation and control of quantum states has been
the subject of considerable work in many other areas of physics.
For example, it is now possible to precisely carve the electronic
wave function of atoms or to excite vibrational modes of molecules
\cite{Wein99,Olsh00,Juds92}, using precisely engineered optical
pulses. Similar techniques have lead to spectacular advances in
nonlinear optics.

A common tool to many of these developments is the use of genetic
algorithms. These multi-dimensional optimization techniques
proceed by parametrizing a control function in terms of a finite
set of coefficients, or ``genes'', a particular set of genes being
called a ``chromosome.'' The genetic algorithm operates on a set
of chromosomes, the ``population.'' Its success in achieving a
design goal is quantified by a ``fitness function,'' a measure of
how close the action of a particular chromosome is to the desired
state. The algorithm proceeds by replacing an ill-fitted fraction
of the population by new chromosomes, the ``offsprings'', that
result from the mating of two parent chromosomes according to some
set of rules; see e.g. Refs. \cite{Davi91,Cole99} for details. In
addition to that controlled combination of chromosomes, random
mutations on single genes prevent the algorithm from getting
trapped in local extrema. The process is iterated until one
chromosome reaches a prescribed fitness value.

Because of collisions, the dynamics of atomic BEC is intrinsically
nonlinear. Hence, it is difficult in general to precisely predict
the effects of control fields on the condensate properties. Indeed,
condensates are emerging as excellent test systems to study
quantum chaos \cite{Carr01,Miln01}. This is a clear indication
that straightforward analytical tools are generally unlikely to be
sufficient in the coherent manipulation and control of BEC. It is
therefore natural to turn instead to the use of genetic
algorithms. The main goal of this paper is to illustrate how they
can be applied to the design of specific momentum states of
condensates.

We specifically consider two examples, the acceleration of a
condensate, and the preparation of a BEC in a coherent
superposition of two momentum states with a prescribed phase
difference. The external control is provided by two
counterpropagating laser fields of adjustable frequency that
provide a time-dependent optical lattice interacting with the
condensate via Bragg scattering. This is a natural choice, since
Bragg scattering is a well-established tool of atom optics: It has
been used in many applications such as the determination of the
coherence properties of condensates \cite{Sten99,Hagl99}, the
implementation of Mach--Zehnder interferometers to image the
condensate phase \cite{Sims00}, the splitting of condensates
\cite{Kozu99}, and the creation of initial states appropriate for
nonlinear mixing processes \cite{Deng99}. Optical
lattices have also been used to investigate physical effects such
as  atomic Landau--Zener tunneling\cite{Niu96,Bhar97}, Bloch
oscillations \cite{Daha96,Peik97a,Peik97b}, and the acceleration
of BECs \cite{Mors01}. Also, a Josephson junction array was
experimentally realized with a BEC in an optical standing wave 
\cite{Ande98,Ande99,Cata01}.

The paper is organized as follows: Section \ref{sec:model} discusses the
model, establishes our notation, and introduces the
``quasi-modes'' of the condensate used in the subsequent analysis.
Section \ref{sec:ga} briefly reviews important aspects of genetic
algorithms. The main results are presented in section \ref{sec:results}, which
illustrates the usefulness of genetic algorithms in condensate
acceleration and in the preparation of macroscopically separated
momentum states of prescribed relative phase. Finally, section \ref{sec:summary}
is a summary and conclusion. Appendix \ref{sec:appendix} gives further
details of the genetic algorithm.

\section{Model}
\label{sec:model}

We consider a condensate consisting of $N$ atoms of mass $M$ at
temperature $T=0$, and placed in a frequency-chirped optical
lattice formed by two counterpropagating laser beams of wave
vector $k_L$ and central frequency $\omega_0$. The electric field 
is then given by
\begin{equation}
E(z,t)
=
\frac{{\cal E}_0}{2}
\left[
e^{i(k_L z-(\omega_0+\delta(t))t)}
+ e^{-i(k_L z+\omega_0 t)}
+ c.c. 
\right],
\label{equ:efield}
\end{equation}
where $\delta(t)$ is a time--dependent frequency difference between
the two beams. We define the detuning $\Delta$ from the atomic resonance 
as $\Delta = \omega_a - \omega_0$, with $\omega_a$ being the 
atomic transition frequency. In order to avoid spontaneous emission we
use far--off resonance light, so that $|\Delta| \gg \Gamma$,
with $\Gamma$ the spontaneous decay rate of the excited atomic level.
We assume that the coupling of the light field from
Eq. (\ref{equ:efield}) to the atomic system is characterized by a Rabi
frequency $\Omega = d{\cal E}_0/\hbar$, where $d$ is the dipole matrix
element of the transition. If the Rabi frequency is small compared to
the detuning, $|\Omega| \ll |\Delta|$,
we can adiabatically eliminate the excited atomic level, which evolves
on a much faster time scale than the lower level. Making use of the
rotating wave approximation which neglects terms varying at twice the
optical frequency $\omega_0$, the resulting time-dependent optical
potential for the lower atomic level is then given by
\begin{equation}
\label{equ:potential}
    V(z,t)= V_0 \cos[2k_L z-\delta(t)t],
\end{equation}
where $V_0 = \hbar |\Omega|^2/2\Delta$ is the lattice depth 
(see also Ref. \cite{Cohe92} for details). 
In order not to violate the adiabatic approximation, the detuning 
$\delta(t)$ must remain small compared to the
detuning $\Delta$, and vary slow on the fast time scale:
\begin{equation}
|\delta| \ll |\Delta|, \quad \left|\frac{d\delta}{dt}\right| \ll \Delta^2.
\end{equation}

The instantaneous phase velocity of the lattice fringes in Eq. (\ref{equ:potential}),
\begin{equation}
 v_{{\rm lat}}(t) \approx \left ( \frac{1}{2k_L} \right )
 \frac{\partial [ \delta(t) t ]}{\partial t},
\end{equation}
must remain much smaller than the speed of light at all times in order
to neglect any time--dependence of the wavenumber $k_L$, as discussed in detail in Ref. 
\cite{Pott01}. 

The control of the state of the BEC is achieved by imposing a time
dependence on the detuning $\delta(t)$ between the
counterpropagating laser beams. It is this time dependence that is
to be determined by the genetic algorithm.

We assume that the condensate is tightly confined in the
transverse direction,  
so that the system can be described semiclassically by a
one-dimensional time-dependent Gross-Pitaevskii equation (GPE).
The condensate evolution in the optical lattice of Eq. (\ref{equ:potential}) is then given by
\begin{eqnarray}
\label{equ:gpe} i\hbar\frac{\partial}{\partial t}\psi(z,t) & = &
\left[ -\frac{\hbar^{2}}{2M}\frac{\partial^{2}}{\partial z^{2}}
+V(z,t) \right ] \psi (z,t)\nonumber \\ &+&
NU_0|\psi(z,t)|^2\psi(z,t),
\end{eqnarray}
where $\psi(z,t)$ is the condensate wave function, 
$U_{0}=4\pi\hbar^2 a/M$, and $a$, the $s$-wave scattering length reduced to one
dimension, is taken to be positive. In the presence of a periodic
potential, one needs to be aware of the possible fragmentation of the 
condensate via a Mott insulator transition. Ref. \cite{Jaks98}
investigates this problem, predicting the onset of fragmentation when the
lattice potential depths are in excess of 5-10 lattice recoil
energies. Here we consider only situations where the lattice
potential is shallow enough that such fragmentation does not occur.
Superfluid effects can likewise be ignored, since the critical
velocity for typical Bose--Einstein
condensates was recently experimentally determined to lie in the $\mbox{mm/s}$
regime \cite{Rama99,Onof00,Burg01} and as we show shortly, Bragg
diffraction only populates momentum side modes of the condensate, 
spaced by momenta $2\hbar k_L$, 
corresponding to velocities far exceeding the critial velocity.  

Our goal in this paper is to manipulate the momentum of the
condensate. It is therefore convenient to introduce the momentum
space condensate wave function
\begin{equation}
\label{equ:fourier} \phi(k,t)
=
\frac{1}{\sqrt{2\pi}}\int\limits_{-\infty}^{\infty} {d}z\,
\psi(z,t) e^{-ikz}.
\end{equation}
Substituting these into Eq. (\ref{equ:gpe}) we obtain the 
corresponding coupled difference-differential
GPEs
\begin{eqnarray}
\label{equ:kspaceGPE} & &i\hbar \frac{\partial}{\partial t}
\phi(k, t) = \frac{\hbar^2 k^2}{2M}\phi(k,t) \nonumber \\ &+&
\frac{V_0}{2} \left[ \phi(k-2k_L)e^{-i\delta(t)t} +
\phi(k+2k_L)e^{i\delta(t)t} \right] \nonumber \\ &+&
\frac{NU_0}{2 \pi} \int dk_1dk_2
\phi(k-k_1+k_2,t)\phi(k_1,t)\phi^\star(k_2,t).
\end{eqnarray}

The spatial extent of the condensates that we have in mind is
large compared to the lattice period $\pi/k_L$. Their initial
momentum distribution is therefore much narrower than $k_L$,
$\Delta k \ll k_L$. From Eq. (\ref{equ:kspaceGPE}), we observe
that the optical lattice couples states separated in momentum by
$k=\pm 2k_L$, and hence leads to a momentum distribution
consisting in general of a ``comb'' of narrow peaks of width
$\Delta k$. Ground state collisions lead to a broadening of these
peaks, but for small enough particle numbers $N$, it can be
expected that this broadening remains small compared to $2 k_L$.
This suggests that it is useful to expand the momentum space condensate
wave function on a basis of ``quasi-modes'' described by the
orthonormal mode functions
\begin{equation}
u_n(k) = \frac{1}{2k_L} \left \{\Theta
[(2n-1)k_L]-\Theta[(2n+1)k_L] \right \}
\end{equation}
as
\begin{equation}
\phi(k,t) = \sum_n \zeta_n(t) u_n(k),
\end{equation}
where
\begin{equation}
\zeta_n(t) = \int_{-\infty}^{\infty} {d}k\, u_n(k) \phi(k,t)=
\int_{(2n-1)k_L}^{(2n+1)k_L} {d}k\, \phi(k,t)
\end{equation}
and the step function $\Theta(x)$ is defined as
\begin{equation}
\Theta(x) = \left\{
\begin{array}
{r@{\mbox{ }:\mbox{ }}l} 0 & x < 0 \\ 
1 & x\ge 0  
\\
\end{array}
\right.
\label{equ:stepfunction}
\end{equation}
The associated ``quasi-mode'' populations $p_n(t)$
are accordingly
\begin{equation}
p_n(t) =|\zeta_n(t)|^2 = \int _{(2n-1)k_L}^{(2n+1)k_L} dk
|\phi(k,t)|^2 .
\end{equation}
In the following, we use a genetic algorithm to find a
time-dependent detuning $\delta(t)$ leading to predetermined
values of the probability amplitudes $\zeta_n(t)$. Before
presenting selected results of this study, though, we review for
completeness some important aspects of this optimization
technique.

\section{Genetic algorithm}
\label{sec:ga}

Genetic algorithms proceed by parametrizing a control function in
terms of a finite set of genes, a particular set of genes being
called a chromosome. The genetic algorithm operates on a set of
chromosomes, the population, whose action on the system to be
controlled is quantified in terms of a fitness function. In the
situation at hand, the control is achieved by the time-dependent
detuning $\delta(t)$, expressed by a truncated Fourier series
\begin{equation}
\label{equ:fourierseries}
\delta_i(t)
=
\sum_{\nu=1}^{m}a_{i \nu}\cos(\nu \omega_R t)+b_{i \nu}
\sin( \nu \omega_R t), \quad i=1,\dots,{\cal N},
\end{equation}
where $\omega_R=\hbar k_L^2/M$ is the recoil frequency. Each
detuning $\delta_i$ is encoded in a chromosome $c_i$ consisting of
$n=2m$ genes $g_{ij}$, each gene corresponding to one particular
Fourier coefficient,
\begin{equation}
c_i(g_{i1},\ldots,g_{in}) = c_i(a_{i1},\ldots,a_{im},b_{i1},\ldots,b_{im}).
\end{equation}
The index $i$ labels a specific chromosome, and the size of the chromosome
population is ${\cal N}$.

\begin{figure}
\begin{center}
\includegraphics[width=0.7\columnwidth]{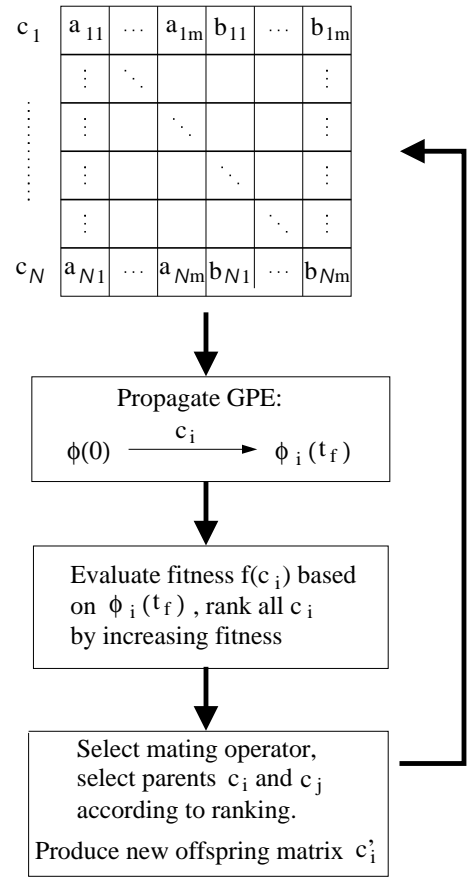}
\vspace{.3 cm} \caption{ Schematics of the genetic algorithm:
In our problem, the fitness is evaluated by
evolving an initial momentum space wavefunction $\phi(0)$ according
to the GPE from Eq. (\ref{equ:gpe}) for a time $t_f$, the dynamics of the
optical lattice being determined by the detuning $\delta_i(t)$. The
final wavefunction is then compared to the optimization goal.
 \label{fig:ga_scheme} }
\end{center}
\end{figure}

Starting from a randomly initialized population,
the genetic algorithm uses a set of mating rules, mutations, and a
problem-specific fitness function $f(c_i)$ to create new
generations of chromosomes, as illustrated in Fig.\ref{fig:ga_scheme}.
The first step consists in selecting parent chromosomes that are
to be combined by mating operators. This is achieved by ranking
the initial ${\cal N}$ chromosomes according to their fitness and
using the so-called ``roulette wheel'' method \cite{Davi91} to
preferentially select parent chromosomes with a high fitness.

In the next step mating operators are selected that generate
a group of ``offspring'' chromosomes $c_i'$ from the ``parent''
population $c_i$. As such, these operators are at the heart of the
genetic algorithm.

Several mating operators may be considered: The
``one-point-crossover'' operator cuts the two parent chromosomes
at a randomly chosen position $\mu$ and swaps the genes according
to
\begin{eqnarray}
& c_1(g_{11},\ldots,g_{1n}) & \nonumber
\\ & c_2(g_{21},\ldots,g_{2n}) &                         \nonumber
\\ & \downarrow &                                        \nonumber
\\ & \mbox{1-point-crossover} &
\\ & \downarrow &                                        \nonumber
\\ & c_1'(g_{11},\ldots,g_{1 \mu},g_{2,\mu+1},\ldots,g_{2n}) & \nonumber
\\ & c_2'(g_{21},\ldots,g_{2 \mu},g_{1,\mu+1},\ldots,g_{1n}) &,
\nonumber \label{equ:xover1}
\end{eqnarray}
with $1\le \mu \le n$. A slightly modified version of this
operator is the ``two-point-crossover'' operator that cuts the two
parent chromosomes at two random positions $\mu_1$ and $\mu_2$ and
then exchanges the genes between these two positions:
\begin{eqnarray}
& c_1(g_{11},\ldots,g_{1n}) & \nonumber
\\ & c_2(g_{21},\ldots,g_{2n}) & \nonumber \\ & \downarrow &
\nonumber
\\ & \mbox{2-point-crossover} &
\\ & \downarrow & \nonumber \\
& c_1'(g_{11},\ldots,g_{1 \mu_1},g_{2,\mu_1+1},\ldots,g_{2,\mu_2-1},
g_{1 \mu_2},\ldots,g_{1n}) & \nonumber \\
& c_2'(g_{21},\ldots,g_{2 \mu _1},g_{1,\mu_1+1},\ldots,g_{1,\mu_2-1},
g_{2 \mu_2},\ldots,g_{2n}), & \nonumber \label{equ:xover2}
\end{eqnarray}
with $1 \le \mu_1 \le \mu_2 \le n$.

Another type of mating operator, the ``average-crossover''
operator, produces just one offspring from the two parent
chromosomes by averaging the genes between two randomly chosen
positions $\mu_1$ and $\mu_2$:
\begin{eqnarray}
& c_1(g_{11},\ldots,g_{1n}) & \nonumber \\
& c_2(g_{21},\ldots,g_{2n}) & \nonumber \\ & \downarrow &
\nonumber
\\ & \mbox{average-crossover} &
\\ & \downarrow &
\nonumber \\
& c_1'(g_{11},\ldots,g_{1 \mu_1},g'_{\mu_1+1},\ldots,g'_{\mu_2-1},
g_{1 \mu_2},\ldots,g_{1n}), & \nonumber
\label{equ:xoveraverage}
\end{eqnarray}
with $1 \le \mu_1 \le \mu_2 \le n$ and
$g'_\kappa = (g_{1 \kappa }+g_{2 \kappa})/2$.

Except for the random location at which the splicing of the
chromosome occurs, the mating algorithms discussed so far are
deterministic. In addition, genetic algorithms also require the
use of ``mutation'' and ``creep.'' These are random operators that
produce one offspring from one parent by altering any gene of the parent
chromosome with a given probability called the mutation rate, respectively
the creep rate:
\begin{eqnarray}
& c(g_1,\ldots,g_n) & \nonumber \\ & \downarrow &
\nonumber
\\ & \mbox{mutation, creep} &
\\ & \downarrow &
\nonumber \\ & c'(g_1,\ldots,g'_\mu,\ldots,g_n).
& \nonumber \label{equ:mutationcreep}
\end{eqnarray}
The mutation operator chooses $g'_\mu$ randomly within some
bounds, whereas the creep operator shifts the old value $g_\mu$ by a
random amount, $g_\mu'=g_\mu+(0.5-r)p_{\rm{creep}}$. Here
$p_{\rm{creep}}$ is a parameter that controls the range of the
shift and $0 \le r \le 1$ a random number.

The operators discussed in this section are the only ones used in
our analysis. When it comes to selecting a specific mating
operator, there are basically two possibilities: The first one
consists of assigning fixed weights to the available operators and
then choosing randomly among them. This is a straightforward
approach, but it suffers from the problem of not discriminating
against mating operators that do not perform well for the
optimization problem at hand. Thus, a second possibility is to
dynamically adjust the operator weights over the course of the
optimization \cite{Davi91,Pear01}. This guarantees that the best
suited operators are applied and allows one to test the
performance of new mating operators. This is done by assigning an
adjustable ``operator fitness'' to each mating operator under
consideration. As such, the mating operators are selected
according to their fitness  the same way as the parent chromosomes
are picked. The details of the procedure used in our simulations
are discussed in the next section and also in the appendix.

\section{Results}
\label{sec:results}

\subsection{Coherent acceleration}
\label{subsec:acceleration}

We now apply the genetic algorithm approach to the manipulation of
the state of a BEC in a chirped optical lattice. As a first
example, we consider the coherent transfer of a condensate
population between the adjacent quasi-modes $u_0$ and $u_1$. Our
motivation here is the need to find efficient ways to accelerate
condensates for atom optics applications. This problem was
theoretically analyzed in Refs. \cite{Choi99,Pott01} and
experimentally demonstrated in Ref. \cite{Mors01} for the case of
a linear dependence of $\delta(t)$, $\delta(t) = \eta t$. While
this chirping of the lattice detuning does lead to a large
mean acceleration of the condensate, it unfortunately leaves a
substantial fraction of the condensate in lower quasi-modes. This
translates into a loss in coherence that is unacceptable, e.g.,
in interferometric applications. The question, then, is whether an
optimized time-dependent detuning determined by a genetic
algorithm can eliminate this problem.

For the sake of illustration, we assume that the condensate is
initially in the zero-momentum quasi-mode, $p_0(t=0)=1$, and seek
a time-dependent detuning such that $p_1(t_f)=1$ after some
predetermined time $t_f$. In that case, the algorithm fitness has
the simple form
\begin{equation}
f(c_i) = p_1(t_f), \label{equ:pop_fitness}
\end{equation}
with an optimal value of unity.

\begin{figure}
\begin{center}
\includegraphics[width=0.85\columnwidth]{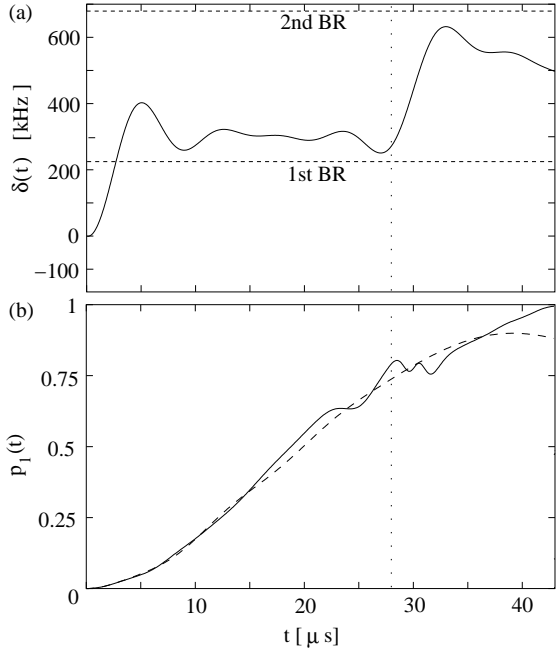}
\vspace{.3 cm} \caption{ Momentum transfer of a Sodium condensate
with a Gaussian spatial distribution of longitudinal width $50
\mbox{ }\mu\mbox{m}$ and transverse width $5 \mbox{ }\mu\mbox{m}$;
$\lambda_L = 985 \mbox{ nm}$ and $V_0=\hbar \omega_R$, $M=3.82\cdot
10^{-26} \mbox{ kg}$ and $a_s=4.9 \mbox{ nm}$. (a) The solid line
is the optimized time-dependent detuning. The dotted lines label
the first and second Bragg resonance (BR) for reference. (b)
Temporal evolution of the mode population $p_1$ for $N=2\cdot
10^6$ atoms: On resonance (dashed) and optimized (solid). The
vertical dotted line denotes the time it takes to resonantly
transfer all population to the quasi-mode $u_1$ in a linear
two-mode system. \label{fig:pop_figure} }
\end{center}
\end{figure}

Fig. \ref{fig:pop_figure} summarizes the results of the
optimization procedure. It compares the optimized population
transfer of a fairly large condensate consisting of
$10^6$ atoms to the case of the same condensate subject to
on--resonance Bragg scattering. In this example, the genetic
algorithm involved a one-point-crossover, a two-point-crossover and an average
cross-over mating operator. In addition, it included two mutation
operators with mutation rates of 0.8 and 0.4, and two creep
operators, both at a rate of 0.9, but different creep parameters,
a ``coarse'' creep with $p_{\rm{creep}} = 0.01$ and a ``finer''
creep operator with $p_{\rm{creep}} = 0.001$. More details of the
simulations are given in the appendix.
While it can be expected that resonant Bragg scattering at the
appropriate frequency transfers perfectly the population of a
small condensate from mode $u_0$ to mode $u_1$, such is not the
case for the large condensate we investigated (dashed line in
Fig.\ref{fig:pop_figure}). In this case, the mean-field
nonlinearity of the condensate  is no longer negligible. It
dynamically shifts the Bragg resonance \cite{Blak00} so that the
transfer efficiency drops to barely over $90\%$ and the maximum
transfer occurs later in time. It is in such nontrivial situations
that genetic algorithms are expected to be useful. Indeed, the
optimal time-dependent detuning $\delta(t)$ found by the genetic
algorithm is highly non-trivial. The temporal dependence of the
detuning $\delta(t)$, which transfers more than $99\%$ of the
population to the quasi-mode $u_1$, reveals that while it is
initially advantageous to remain close to the Bragg resonance
frequency, as indicated by the rather flat portion of the
detuning, it eventually becomes necessary to drastically couple
the condensate atoms to higher momentum modes so as to drag the
remaining population to the final state $u_1$ \cite{foot1}.

\begin{figure}
\begin{center}
\includegraphics[width=0.85\columnwidth]{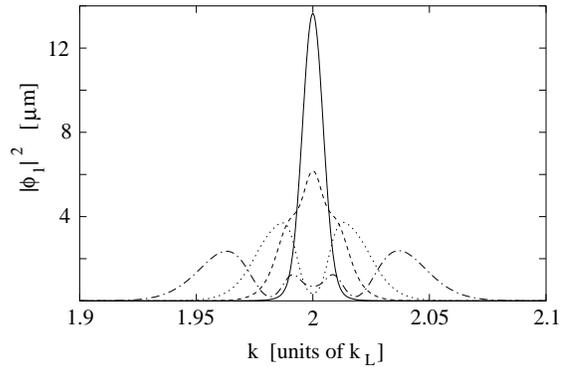}
\vspace{.3 cm} \caption{ Effect of the nonlinearity on the
momentum space wave function: Momentum space densities within
quasimode $u_1$ after optimizing the transfer, so that $p_1 >
0.99$ after time $t_f = 43 \mbox{ }\mu\mbox{s}$; $N=1\cdot 10^5$
(solid), $N=5\cdot 10^5$ (dashed), $N=1\cdot 10^6$ atoms (dotted),
and $N=2\cdot 10^6$ (dashed--dotted). \label{fig:pop_number}}
\end{center}
\end{figure}

The effect of the mean-field energy is further illustrated in Fig.
\ref{fig:pop_number}, which shows the final momentum distribution
$\phi_1(k)$ within the quasi-mode $u_1$ for various numbers of
atoms in the condensate for optimal transfer. While this
distribution remains extremely narrow compared to the quasi-mode
width $2 k_L$, collisions lead to a substantial reshaping and
broadening within that mode.

\subsection{Coherent superposition}
\label{subsec:superposition}

In a second application,  we set out to design an equal-weight
superposition of the two quasi-mode states $u_0$ and $u_1$
\begin{equation}
\phi(k,t_f)
=
\frac{1}{\sqrt{2}} \left(u_0 e^{i\varphi_0} + u_1
e^{i\varphi_1} \right),
\end{equation}
with a prescribed relative phase $\Delta \varphi=
\varphi_1-\varphi_0$. In contrast to the previous example, we now
want to control two properties of the quantum state, the relative
phase as well as the population in each state. The fitness
function to be optimized is therefore more complicated.

\begin{figure}
\begin{center}
\includegraphics[width=0.9\columnwidth]{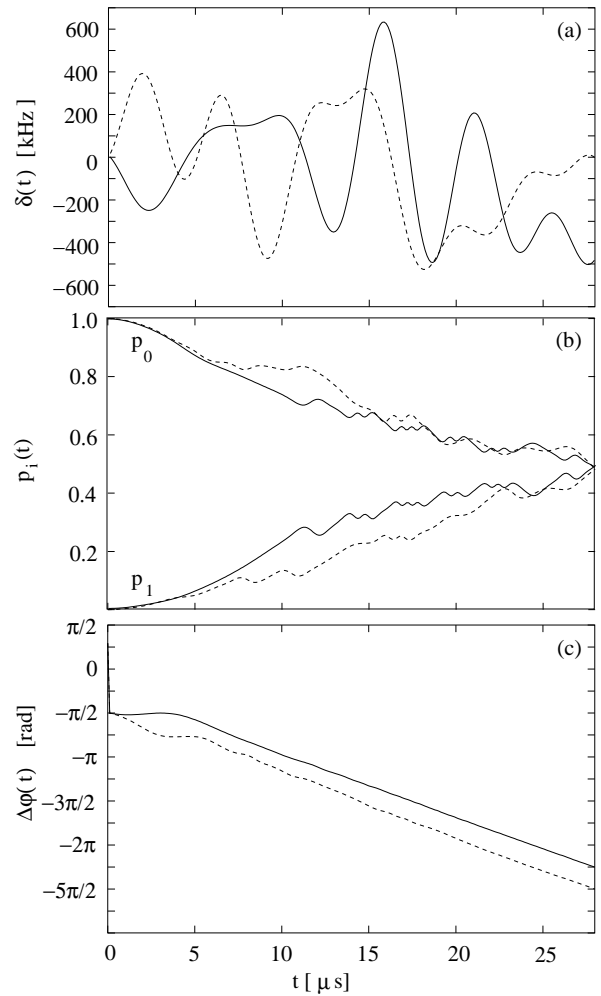}
\vspace{.3 cm} \caption{ Excitation of an equal-weight coherent
superposition of momentum states with relative phases,
$\Delta\varphi=-\pi/4$ (solid) and $\Delta\varphi=-\pi/2$
(dashed). The parameters are the same as in Fig.
\ref{fig:pop_figure}, except that the condensate has a Gaussian
spatial distribution of longitudinal width $100 \mbox{
}\mu\mbox{m}$ and $N=5\cdot 10^4$. (a) Optimized time--dependent
detuning used to create the superposition. (b) Temporal evolution
of the mode population of the two involved modes. (c) Temporal
evolution of relative phase of the two quasi-modes at the center
of their momentum distribution, see Fig. 5.
\label{fig:phase_figure} }
\end{center}
\end{figure}

We choose to employ a so-called ``penalty function'' $P(c_i)$ for
the optimization of the quasi-mode populations \cite{Cole99}. The
goal of $P(c_i)$ is to decrease the fitness of chromosomes that do
not fulfill the desired requirements, thereby steering the
population towards the target values. A prototype penalty function
is
\begin{equation}
P(c_i) = \left\{
\begin{array}
{r@{\mbox{ }:\mbox{ }}l} 1.5 & p_0(c_i), p_1(c_i)
> 0.465 \\ 1.0 & p_0(c_i), p_1(c_i) > 0.47  \\ 0
& p_0(c_i), p_1(c_i) > 0.475 \\ 100 & \mbox{else}
\\
\end{array}
\right.
\label{equ:penalty}
\end{equation}
The fitness function for this optimization problem is then given by
\begin{equation}
f(c_i)
=
1-\left|\alpha(c_i)-\Delta\varphi\right|-P(c_i), \label{equ:phase_fitness}
\end{equation}
where, as we recall, $c_i$ corresponds to a specific realization of
the time-dependent detuning $\delta(t)$ and $\alpha(c_i)$ is the
relative phase corresponding to this realization. This fitness function
reaches its maximum, unity in this case, when the populations are within the specified
range and the phase difference is exactly as prescribed. The
results for the optimization for the two cases $\Delta
\varphi=-\pi/2$ and $\Delta \varphi= -\pi/4$ are shown in Fig.
\ref{fig:phase_figure}. For the genetic algorithm we used the same
operators and their parameters as in Sec. \ref{subsec:acceleration}.
More details of the simulations are given in the appendix.

Fig. \ref{fig:phase_kspace} shows the momentum distributions
$\phi_0(k)$ and $\phi_1(k)$ of the
condensate within each of the two quasi-modes $u_0$ and $u_1$, as
well as the corresponding phases for the case
$\Delta\varphi=-\pi/2$. Clearly, the genetic algorithm converges
to the stated goal, and produces a condensate in the desired
coherent superposition.

In particular, we observe that the relative phase of the two
components is approximately constant in the region where the
condensate wave function is different from zero. The optimization
goal of $\Delta\varphi=-\pi/2$ is achieved with an accuracy of
over $99\%$ at the center of the mode. The curvature of the phase
at the wings of the wave function is due to nonlinear phase shifts
accumulated during the Hamiltonian evolution. It could be reduced
by decreasing the number of atoms in the condensate.
\begin{figure}
\begin{center}
\includegraphics[width=0.95\columnwidth]{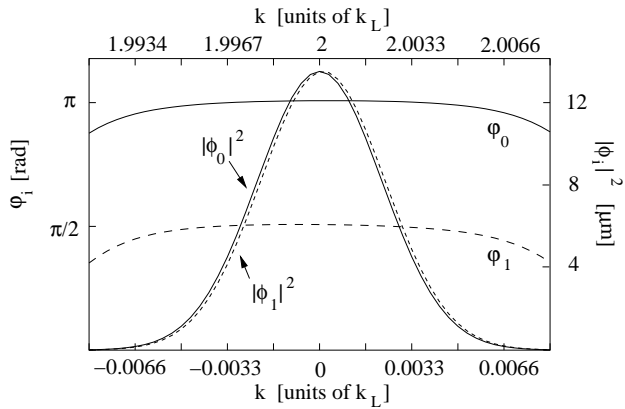}
\vspace{.3 cm} \caption{ Density profiles and phases of the two
momentum modes after the optimization at $t=t_f$.
The lower abscissa corresponds to the momentum
wavefunction $\phi_0$ with phase $\varphi_0$ centered around
$k=0$, the upper to $\phi_1$ with phase $\varphi_1$ centered
around $k=2k_L$. Note that $p_1(t_f),p_2(t_f) \ge 0.475$.
\label{fig:phase_kspace} }
\end{center}
\end{figure}

\section{Summary and conclusion}
\label{sec:summary}

In summary, we have demonstrated the usefulness of genetic
algorithms in the control and manipulation of the quantum states
of Bose-Einstein condensates. This was illustrated in two
examples, the coherent population transfer between momentum states
in a large condensate where the mean-field effects are important,
and the creation of coherent superpositions of states of
prescribed population and relative phase. We found that
time--dependent Bragg scattering combined with the powerful
optimization capabilities of genetic algorithms, provides a
novel tool for quantum state design and coherent control in
linear and nonlinear atom optics. The extension of these ideas to
integrated atom optics appears particularly promising. Future
theoretical work will generalize these concepts to systems with
more degrees of freedom, such as e.g. multi-component condensates,
and to additional control mechanisms such as time-dependent
magnetic fields. The extension of this work to quantum-degenerate
Bose-Fermi mixtures also appears promising.

\acknowledgements We thank J.~V. Moloney for CPU time. This work
is supported in part by the U.S.\ Office of Naval Research under
Contract No.\ 14-91-J1205, by the National Science Foundation
under Grant No.\ PHY98-01099, by the NASA Microgravity Program
Grant NAG8-1775, by the U.S.\ Army Research Office, and by the
Joint Services Optics Program.

\appendix
\section{Details of the genetic Algorithm}
\label{sec:appendix}

In this Appendix we give specific details of the implementation of
the genetic algorithm in the problem of BEC state engineering in
an optical lattice.

\begin{figure}
\begin{center}
\includegraphics[width=0.99\columnwidth]{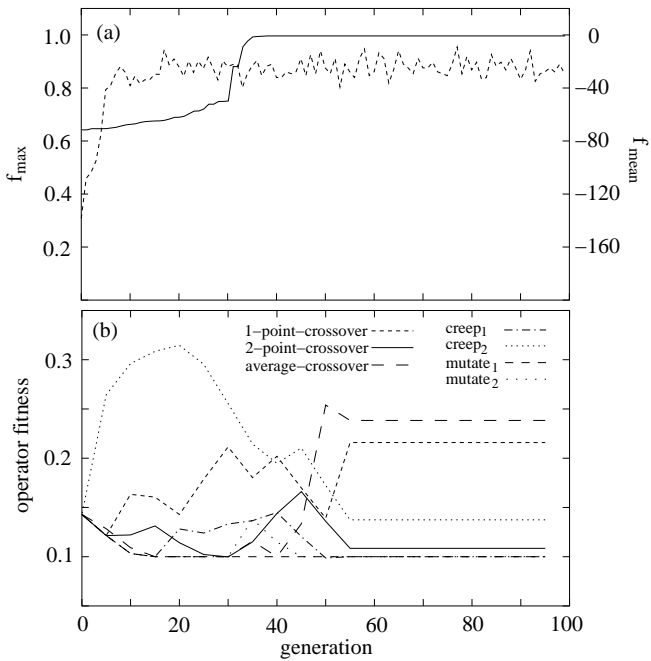}
\vspace{.3 cm} \caption{ A typical run of the genetic algorithm
for the superposition state engineering: (a) Maximum fitness
(solid) and mean fitness (dashed) of the population as a function
of the generation. We started the optimization with pre--optimized
chromosomes from previous simulations, which explains the high
initial maximum fitness of over $60\%$. (b) Operator fitness as a
function of the generation, starting with equal fitness for all
operators. The minimum operator fitness is set equal to $0.1$ for all
operators. \label{fig:ga_figure} }
\end{center}
\end{figure}

Over the course of the optimization we monitor the maximum fitness
$f_{\rm{max}}$ of a population,
\begin{equation}
f_{\rm{max}} = \max\{f(c_i) : i=1,\ldots,{\cal N}\}.
\end{equation}
Obviously, if at least the best chromosome is kept from the old
generation, the maximum fitness is a monotonically increasing
function. This feature, called ``elitism'', is used throughout our
simulations. Another observable of interest is the mean fitness
$f_{\rm{mean}}$ of a population,
\begin{equation}
f_{\rm{mean}} = \frac{1}{{\cal N}} \sum^{\cal N}_{i=1} f(c_i).
\end{equation}
A typical evolution of these two quantities is shown in Fig.
\ref{fig:ga_figure}(a). The maximum fitness increases
monotonically to reach a value close to the optimum after about 40
generations. In contrast, the mean fitness rises over the course
of the first 10 generations and then exhibits fluctuations due to
the stochastic character of the genetic algorithm: Parts of the
population are replaced by randomly created chromosomes from one
generation to the next and randomize the mean fitness value. In
our simulations we used populations of size ${\cal N}=50-100$ and
performed the optimization over $50-100$ generations. We always
kept the best chromosomes of a generation and replaced $80\%-90\%$
of the population by newly generated chromosomes. For the
population transfer of Sec. \ref{subsec:acceleration} we used 16, and for the
superposition state engineering of Sec. \ref{subsec:superposition} $26$ genes per
chromosome. The gene boundaries were chosen as
$-0.7 \omega_R\le a_{i\nu}, b_{i\nu} \le 0.7 \omega_R$.

As mentioned in Sec. \ref{sec:ga}, we use an adaptive operator
technique, where the operators themselves are dynamically assigned a fitness based on
their performance. Choosing a particular mating operator 
via a roulette--wheel method \cite{Davi91} then assures that good operators 
are employed more often in the mating process. 
If any operator produces an offspring that is better than the best 
chromosome of the previous generation, we reward this operator by giving
it a credit proportional to the increase in fitness it caused.  
Also, we pass half of this given credit back to the operator that created the
parent chromosome involved in producing the better offspring. 
Thus operators that perform well and also their direct ancestors
can accumulate credit over the run of the simulation. Passing credit back to previous
operators enables us to reward pairs of operators that work well together at a
certain stage of the optimization process.
In our simulations we adjust the fitness of all operators every five generations
based on the credit they accumulated during that period: The new operator fitness is then a weighted sum
of the old fitness (the ``basis portion'', in our case $85\%$) and the accumulated credit 
(the ``adaptive portion'', in our case $15\%$). The more credit an operator accumulates the 
higher will be its new fitness and the more likely it is that it will be chosen in future mating processes.
Since the total operator fitness is set constant, we introduce a lower bound of $0.1$ to the operator
fitness, thereby preventing operators that do not perform well over several generations from being practically
expelled from the pool of operators.

Fig. \ref{fig:ga_figure}(b) shows a typical evolution of the
fitness of the individual mating operators in the superposition
state engineering problem from Sec. \ref{subsec:superposition}.
The creep and one-point-crossover operators perform well for the
first 30 generations. For subsequent generations the
two-point-crossover and the average-crossover operators take over
and help increasing the maximum fitness, which approaches its
optimum value (unity in the present example). There is no further
improvement in the remaining 50 generations, the fitness of the
various operators staying constant. The random mutation operators
never perform well in the problem at hand. Consequently their
fitness is quickly reduced to the lower bound. The creep
operators, which are basically controlled mutations, perform much
better and could replace the pure mutation operators.


\begin{references}
\bibitem{foot1}
The relatively fast initial increase of the detuning above the level of the
first Bragg resonance makes up for the fact that at $t=0$ we start
with zero detuning and couple to negative momentum modes.
\end{references}
\end{document}